\documentclass[pra,twocolumn,showpacs]{revtex4-1}
\usepackage{amsmath,amssymb,bm,graphicx,hyperref}
\allowdisplaybreaks[1]

\renewcommand\d{\partial}
\newcommand\grad{\bm{\nabla}}
\newcommand\+{\dagger}
\newcommand\<{\langle}
\renewcommand\>{\rangle}
\newcommand\eps{\epsilon}
\newcommand\A{\bm{A}}
\newcommand\J{{\bm{J}}}
\renewcommand\r{{\bm{r}}}
\renewcommand\Re{\mathrm{Re}}
\renewcommand\Im{\mathrm{Im}}
\newcommand\Tr{\mathrm{Tr}}
\newcommand\vF{v_\mathrm{F}}
\newcommand\sgn{\mathrm{sgn}}
\newcommand\T{\mathcal{T}}
\DeclareMathOperator\arctanh{arctanh}

\begin{document}

\title{Renormalization group analysis of graphene with a supercritical Coulomb impurity}

\author{Yusuke Nishida}
\affiliation{Department of Physics, Tokyo Institute of Technology,
Ookayama, Meguro, Tokyo 152-8551, Japan}

\date{May 2016}

\begin{abstract}
We develop a field-theoretic approach to massless Dirac fermions in a supercritical Coulomb potential.
By introducing an Aharonov--Bohm solenoid at the potential center, the critical Coulomb charge can be made arbitrarily small for one partial-wave sector, where a perturbative renormalization group analysis becomes possible.
We show that a scattering amplitude for reflection of particle at the potential center exhibits the renormalization group limit cycle, i.e., log-periodic revolutions as a function of the scattering energy, revealing the emergence of discrete scale invariance.
This outcome is further incorporated in computing the induced charge and current densities, which turn out to have power-law tails with coefficients log-periodic with respect to the distance from the potential center.
Our findings are consistent with the previous prediction obtained by directly solving the Dirac equation and can in principle be realized by graphene experiments with charged impurities.
\end{abstract}


\maketitle

\section{Introduction}
One of the great successes of quantum mechanics was Bohr's explanation for the stability and structure of atoms~\cite{Bohr:1913}.
Quantum mechanics was later married with special relativity to give birth to relativistic quantum mechanics.
In particular, Paul Dirac derived the celebrated wave equation, which was solved for an electron in a nuclear Coulomb potential leading to the fine structure of atoms~\cite{Dirac:1928}.
However, when the nuclear charge $Z$ exceeds the reciprocal of the fine structure constant $\alpha=e^2/(4\pi\varepsilon_0\hbar c)\approx1/137$, ordinary solutions to the Dirac equation break down by producing a complex spectrum in the lowest angular-momentum channel~\cite{Berestetskii:1982}.
While this problem is usually evaded by allowing a finite size for the charged nucleus~\cite{Pomeranchuk:1945,Zel'dovich:1971}, it is less well known that universal physics independent of such short-distance details can be extracted by reimposing the appropriate boundary condition at the location of the point nucleus~\cite{Case:1950,Gitman:2012}.

To elaborate along the lines of Ref.~\cite{Nishida:2014}, we recall that the radial Dirac equation in a supercritical angular-momentum channel admits two linearly independent solutions which, toward the potential center, behave as powers of the radius $\sim r^{\pm i\gamma}$ with exponents complex conjugate to each other.
Because they correspond to inward and outward spherical waves, the unitarity requires superposing the two solutions equally.
To this end, a new length scale $r_*$ must be introduced to unify their dimensions, for example, by multiplying the one solution $\sim r^{-i\gamma}$ by $r_*^{2i\gamma}$.
Obviously, the resulting physics is invariant under the replacement of $r_*$ with its multiple by a factor of $e^{n\pi/\gamma}$ for arbitrary integer $n$.
The physical consequences of such discrete scale invariance include not only the geometric sequence of so-called atomic collapse resonances~\cite{Pereira:2007,Shytov:2007a,Shytov:2007b} but also the log-periodic oscillation in the vacuum polarization, which are universally related through the single quantity $r_*$~\cite{Nishida:2014}.

While these intriguing phenomena caused by a supercritical Coulomb potential have been predicted by directly solving the Dirac equation~\cite{Pereira:2007,Shytov:2007a,Shytov:2007b,Nishida:2014}, the present paper is aimed at developing an alternative field-theoretic approach to such peculiarities of relativistic quantum mechanics.
In particular, we focus on massless Dirac fermions in two dimensions because they emerge in low-energy physics of graphene with an effective fine structure constant as large as $e^2/(4\pi\varepsilon_0\hbar\vF)\approx2$ and thus just a few charged impurities are sufficient to produce the supercritical Coulomb potential~\cite{Wang:2012,Wang:2013,Luican-Mayer:2014,Mao:2015}.
Furthermore, the critical Coulomb charge in two-dimensional systems can in principle be controlled by introducing an Aharonov--Bohm solenoid at the potential center~\cite{Levitov:2015}, which facilitates our theoretical analysis as discussed below.

\section{Field-theoretic formulation}
The second-quantized Hamiltonian describing massless Dirac fermions in two dimensions is
\begin{align}\label{eq:hamiltonian}
H = \int\!d\r\,\Psi^\+(\r)\{\vF[-i\hbar\grad+e\A(\r)]\cdot\bm\sigma - eV(\r)\}\Psi(\r),
\end{align}
where $-e<0$ is the electron charge but the electron-electron interaction is neglected.
The Coulomb potential produced by a net charge of $Ze$ centered at the origin is
\begin{align}\label{eq:coulomb}
V(\r) = \frac{Ze}{4\pi\varepsilon_0r},
\end{align}
while the Aharonov--Bohm solenoid centered at the same point provides
\begin{align}\label{eq:solenoid}
\A(\r) = \frac{\Phi}{2\pi}\left(-\frac{y}{r^2},\frac{x}{r^2}\right),
\end{align}
with a net magnetic flux of $\Phi$.
This constitutes our Aharonov--Bohm--Coulomb--Dirac (ABCD) problem.

Because of the rotational symmetry in Eq.~(\ref{eq:hamiltonian}), the partial-wave expansion of
\begin{align}\label{eq:expansion}
\Psi(\r) = \sum_{j=-\infty}^\infty
\begin{pmatrix}
e^{i(j-\frac12)\theta} & 0 \\
0 & ie^{i(j+\frac12)\theta}
\end{pmatrix}
e^{-i\frac\pi4\sigma_1}\frac{\psi_j(r)}{\sqrt{2\pi r}}
\end{align}
decouples the Hamiltonian into different partial-wave sectors as
\begin{align}\label{eq:radial}
H &= \hbar\vF\sum_{j=-\infty}^\infty\int_0^\infty\!dr\,\psi_j^\+(r)
\left(-i\d_r\sigma_3 - \frac{g}{r}\sigma_0 + \frac{j+\phi}{r}\sigma_1\right) \notag\\
&\quad \times \psi_j(r).
\end{align}
Here $j=\pm1/2,\pm3/2,\dots$ is the total angular momentum and we defined the dimensionless Coulomb coupling
\begin{align}
g \equiv \frac{Ze^2}{4\pi\varepsilon_0\hbar\vF},
\end{align}
as well as the magnetic flux in units of $h/e$:
\begin{align}
\phi \equiv \frac{e\Phi}{2\pi\hbar}.
\end{align}
Besides the kinetic term, each radial Hamiltonian in Eq.~(\ref{eq:radial}) consists of the Coulomb potential and the centrifugal potential, both of which are in the scale-invariant form of $1/r$.
Because the role of the Aharonov--Bohm solenoid is just to shift the total angular momentum by $\phi$, all solutions to the radial Dirac equation obtained previously in Refs.~\cite{Pereira:2007,Shytov:2007a,Shytov:2007b,Nishida:2014} remain valid by replacing $j$ with $j+\phi$.
In particular, the critical Coulomb coupling becomes $|j+\phi|$ which can be made arbitrarily small, in principle, by controlling the magnetic flux~\cite{Levitov:2015} (see also Ref.~\cite{Chakraborty:2011}).

Physical quantities of our interest are the charge density $\rho(\r)=-e\<\Psi^\+(\r)\Psi(\r)\>$ and the current density $\J(\r)=-e\vF\<\Psi^\+(\r)\bm\sigma\Psi(\r)\>$ in the ground state of the ABCD Hamiltonian (\ref{eq:hamiltonian}).
With the use of the partial-wave expansion (\ref{eq:expansion}), the charge density can be expressed as
\begin{align}\label{eq:charge}
\rho(r) = -\frac{e}{2\pi r}\sum_{j=-\infty}^\infty\<\psi_j^\+(r)\sigma_0\psi_j(r)\>,
\end{align}
and the current density as
\begin{align}
J_r(r) = -\frac{e\vF}{2\pi r}\sum_{j=-\infty}^\infty\<\psi_j^\+(r)\sigma_3\psi_j(r)\>
\end{align}
in the radial direction and as
\begin{align}\label{eq:current}
J_\theta(r) = -\frac{e\vF}{2\pi r}\sum_{j=-\infty}^\infty\<\psi_j^\+(r)\sigma_1\psi_j(r)\>
\end{align}
in the angular direction.
Therefore, our task is to compute their contributions from each partial-wave sector.

\begin{figure*}[t]
\includegraphics[width=0.92\textwidth]{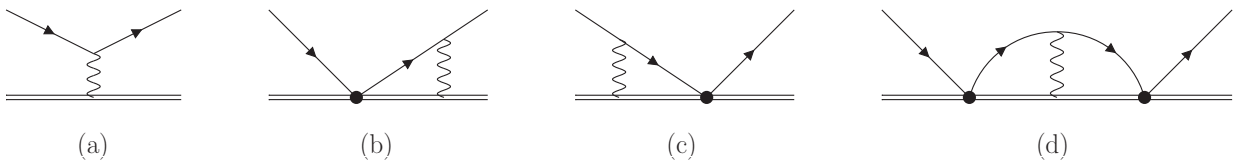}
\caption{\label{fig:renormalization}
Feynman diagrams to the lowest order in $g_0$ and $g_1$ that renormalize $v_{0,1,2,3}$.
The solid lines represent the fermion propagator in Eq.~(\ref{eq:propagator}), the wavy lines represent the $1/r$ potentials in Eq.~(\ref{eq:potential}), the dots represent the contact terms in Eq.~(\ref{eq:contact}), and the double lines represent the potential center located at $r=0$.}
\end{figure*}

To develop a field-theoretic approach to a particular partial-wave sector, we recall that the point charge (\ref{eq:coulomb}) and the line solenoid (\ref{eq:solenoid}) assumed implicitly above are effective descriptions valid at distances sufficiently longer than the actual charge and solenoid radii $\sim\Lambda^{-1}$.
This motivates us to consider a semi-infinite one-dimensional problem defined by the following imaginary-time action:
\begin{subequations}\label{eq:action}
\begin{align}
S &= \int_{-\infty}^\infty\!d\tau\int_0^\infty\!dr\,\psi_j^\+(\tau,r)(\d_\tau\sigma_0-i\d_r\sigma_3)\psi_j(\tau,r) \label{eq:kinetic}\\
& - \int_{-\infty}^\infty\!d\tau\int_{\Lambda^{-1}}^\infty\!\frac{dr}{r}\,\psi_j^\+(\tau,r)(g_0\sigma_0+g_1\sigma_1)\psi_j(\tau,r) \label{eq:potential}\\
& - \sum_{a=0}^3\int_{-\infty}^\infty\!d\tau\,\psi_j^\+(\tau,0)v_a\sigma_a\psi_j(\tau,0). \label{eq:contact}
\end{align}
\end{subequations}
Here we set $\hbar=\vF=1$ and dimensionless couplings for the $1/r$ potentials in Eq.~(\ref{eq:potential}) are $g_0\equiv g$ and $g_1\equiv-(j+\phi)$ according to Eq.~(\ref{eq:radial}), which are now cut off at $r=\Lambda^{-1}>0$.
On the other hand, new contact terms with dimensionless couplings $v_{0,1,2,3}$ are introduced at the potential center to ensure the cutoff independence of long-distance physics.

The physical meaning of Eq.~(\ref{eq:contact}) can be further clarified by writing down the bare propagator generated by Eq.~(\ref{eq:kinetic}), which is
\begin{align}\label{eq:propagator}
G(\eps,r-r') &\equiv \int_{-\infty}^\infty\!d\tau\,e^{i\eps\tau}\<\T\psi_j(\tau,r)\psi_j^\+(0,r')\>_0 \notag\\
&= \frac{\sgn(\eps)\sigma_0+\sgn(r-r')\sigma_3}{-2i}e^{-|\eps||r-r'|},
\end{align}
and thus nonzero only in the upper-left (lower-right) element when $\sgn(\eps)\sgn(r-r')>0$ ($<0$).
Therefore, the upper and lower components of $\psi_j$ for $\eps>0$ correspond to the outward ($r>r'$) and inward ($r<r'$) spherical waves, respectively, while their roles are reversed for $\eps<0$.
This in turn means that the upper-right (lower-left) element of $\sum_{a=0}^3v_a\sigma_a$ in Eq.~(\ref{eq:contact}) plays the role of reflecting particle at the potential center for $\eps>0$ ($<0$), while the other elements play no physical role in our semi-infinite one-dimensional problem defined only for $r,r'>0$.
Accordingly, the corresponding coupling $v_1-i\,\sgn(\eps)v_2$ represents a scattering amplitude for reflection of particle at the potential center and its energy dependence is to be determined by the renormalization group analysis.

\section{Renormalization group analysis}
\begin{figure}[b]
\includegraphics[width=0.98\columnwidth]{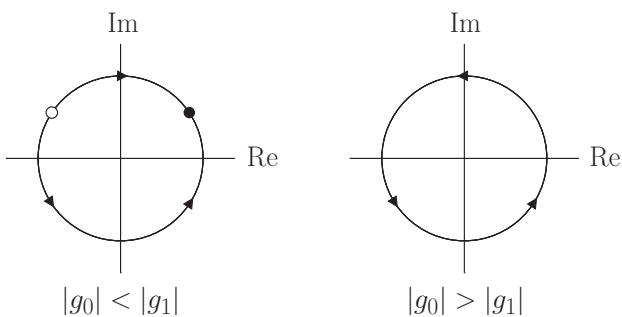}
\caption{\label{fig:flow}
Renormalization group flow of $v_1-i_\eps v_2$ in its complex plane.
In the subcritical case (left figure), the running coupling subject to the unitarity condition flows on the unit circle from the ultraviolet fixed point ($\circ$) into the infrared fixed point ($\bullet$).
On the other hand, in the supercritical case (right figure), the two fixed points are ``pair annihilated'' from the unit circle and the renormalization group limit cycle emerges.}
\end{figure}

To facilitate our theoretical analysis, we regard $g_0$ and $g_1$ as small perturbations.
To their lowest order at $O(g)$, there are four distinct diagrams that renormalize $v_{0,1,2,3}$ as depicted in Fig.~\ref{fig:renormalization}.
After straightforward calculations summarized in Appendix~\ref{sec:renormalization}, the renormalization group equation that governs the running of $v_1-i_\eps v_2$ is found to be
\begin{align}\label{eq:running}
\frac{d(v_1-i_\eps v_2)}{d\ln\Lambda} = g_1+ 2i_\eps g_0(v_1-i_\eps v_2) - g_1(v_1-i_\eps v_2)^2,
\end{align}
with $i_\eps\equiv i\,\sgn(\eps)$.
We note that the beta function is quadratic in terms of $v_1-i_\eps v_2$ which is not altered even by higher-order corrections in $g_0$ and $g_1$.
Furthermore, because the complex coupling $v_1-i_\eps v_2$ represents a scattering amplitude for reflection of particle at the potential center, the unitarity requires its modulus to be unity.
The resulting solution exhibits qualitatively different behavior depending on whether the Coulomb coupling is subcritical or supercritical as illustrated in Fig.~\ref{fig:flow} (see also Refs.~\cite{Kaplan:2009,Moroz:2010}).

In the case of subcritical Coulomb coupling $|g_0|<|g_1|$, the general solution to the renormalization group equation (\ref{eq:running}) subject to the unitarity condition $|v_1-i_\eps v_2|=1$ can be obtained as
\begin{align}\label{eq:subcritical}
v_1-i_\eps v_2 = i_\eps\frac{g_0}{g_1} + \frac{\bar\gamma}{g_1}\tanh\!\left(\bar\gamma\ln\frac{\Lambda}{2|\eps|}+\arctanh\frac{i_\eps\bar\gamma}{g_0\pm g_1}\right),
\end{align}
with $\bar\gamma\equiv\sqrt{g_1^2-g_0^2}$.
Here an arbitrary energy scale $\eps$ should be identified with the scattering energy of particle up to an unimportant factor because it is the only energy scale possible in our scattering amplitude.
Accordingly, we find that the resulting scattering amplitude flows on the unit circle into an infrared fixed point at
\begin{align}\label{eq:infrared}
\lim_{|\eps|/\Lambda\to0}v_1-i_\eps v_2 \to i_\eps\frac{g_0}{g_1} + \frac{\bar\gamma}{g_1}
\end{align}
in the low-energy or large-cutoff limit.

On the other hand, in the case of supercritical Coulomb coupling $|g_0|>|g_1|$, the general solution to the renormalization group equation (\ref{eq:running}) turns into
\begin{align}\label{eq:supercritical}
v_1-i_\eps v_2 = i_\eps\frac{g_0}{g_1} - \frac{\gamma}{g_1}\tan\!\left(\gamma\ln\frac{\Lambda}{2|\eps|}+\arctan\frac{i_\eps\gamma}{g_0+g_1}\right)
\end{align}
by replacing $\bar\gamma$ in Eq.~(\ref{eq:subcritical}) with $i\gamma\equiv i\sqrt{g_0^2-g_1^2}$.
Here the upper sign in $g_0\pm g_1$ was chosen without loss of generality because their difference can be absorbed by the redefinition of $\Lambda$.
Remarkably, we find that the resulting scattering amplitude subject to the unitarity condition $|v_1-i_\eps v_2|=1$ exhibits log-periodic revolutions in its complex plane as a function of the scattering energy $\eps$.
This is nothing short of the renormalization group limit cycle revealing the emergence of discrete scale invariance by a factor of $e^{n\pi/\gamma}$ for arbitrary integer $n$~\cite{Wilson:1971}.
In particular, when $\eps>0$ is analytically continued to a complex variable $iE$, the scattering amplitude (\ref{eq:supercritical}) has an infinite tower of poles at
\begin{align}\label{eq:resonance}
E_n = -\frac{i}{2}\Lambda_*e^{-(\frac12+n)\pi/\gamma},
\end{align}
with
\begin{align}
\Lambda_* \equiv \Lambda\exp\!\left(\frac1\gamma\arctan\frac{i\gamma}{g_0+g_1}\right),
\end{align}
which corresponds to the geometric sequence of atomic collapse resonances~\cite{Pereira:2007,Shytov:2007a,Shytov:2007b}.
We note that the renormalization group limit cycle in the context of graphene with a supercritical Coulomb impurity was also discussed in Refs.~\cite{Gorsky:2014,Bulycheva:2014} from a different perspective.

\section{Induced charge and current}
We now study the physical consequences of our findings from the renormalization group analysis to the charge and current densities induced by the Coulomb potential with the Aharonov--Bohm solenoid.
Feynman diagrams that potentially contribute to $\<\psi_j^\+(r)\sigma_{0,3,1}\psi_j(r)\>$ at $O(1)$ and $O(g)$ are depicted in Figs.~\ref{fig:polarization} and \ref{fig:renormalization}, respectively, whose expressions are summarized in Appendix~\ref{sec:polarization}.

\begin{figure}[t]
\includegraphics[width=0.4\columnwidth]{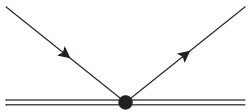}
\caption{\label{fig:polarization}
Feynman diagram at $O(1)$ that potentially contributes to the charge and current densities by closing the two external solid lines at a position $r>0$ with an appropriate $\sigma$ matrix.
The possible contributions at $O(g)$ are already presented in Fig.~\ref{fig:renormalization}.}
\end{figure}

The contributions of Figs.~\ref{fig:polarization} and \ref{fig:renormalization}(d) to the charge density $\<\psi_j^\+(r)\sigma_0\psi_j(r)\>$ vanish and thus the lowest contributions are $O(g)$ and are evaluated as
\begin{align}\label{eq:density}
\mathrm{Fig.\,\ref{fig:renormalization}(a)} = \frac{g_0}{\pi r}
\end{align}
and
\begin{align}
\mathrm{Fig.\,\ref{fig:renormalization}(b)}
&= \mathrm{Fig.\,\ref{fig:renormalization}(c)} \notag\\
&= -\frac{g_1}{\pi}\int_0^\infty\!d\eps\,\Im[v_1-i_\eps v_2]\,\Gamma(0,2\eps r),
\end{align}
where $(v_1-i_\eps v_2)^*=(v_1-i_\eps v_2)|_{\eps\to-\eps}$ is used.
Their sum leads to $\<\psi_j^\+(r)\sigma_0\psi_j(r)\>=0$ in the subcritical case of Eq.~(\ref{eq:infrared}) but
\begin{align}
\<\psi_j^\+(r)\sigma_0\psi_j(r)\>
= \frac{\gamma}{\pi r}\int_0^\infty\!dz\,\Im\!\left[\tan\!\left(\gamma\ln\frac{r\Lambda_*}{z}\right)\right]\Gamma(0,z)
\end{align}
in the supercritical case of Eq.~(\ref{eq:supercritical}).
Therefore, while the subcritical Coulomb potential does not induce a tail in the charge density~\cite{Terekhov:2008}, a power-law tail is induced by the supercritical Coulomb potential with its coefficient being a log-periodic function of $r$.
Because only one partial-wave sector can be supercritical for $|g_0|\ll1$, the charge density (\ref{eq:charge}) including all partial waves is found to be
\begin{align}\label{eq:charge_tail}
\rho(r) = -\frac{e}{2\pi^2r^2}\left\{\gamma\int_0^\infty\!dz\,\Im\!\left[\tan\!\left(\gamma\ln\frac{r\Lambda_*}{z}\right)\right]\Gamma(0,z)\right\},
\end{align}
which is consistent with the previous prediction obtained by directly solving the Dirac equation~\cite{Nishida:2014}.

Similarly, the contributions of Figs.~\ref{fig:polarization} and \ref{fig:renormalization}(a) and \ref{fig:renormalization}(d) to the radial current density $\<\psi_j^\+(r)\sigma_3\psi_j(r)\>$ vanish and the other nonzero contributions are evaluated as
\begin{align}
\mathrm{Fig.\,\ref{fig:renormalization}(b)}
&= -\mathrm{Fig.\,\ref{fig:renormalization}(c)} \notag\\
&= -i\frac{g_1}{\pi}\int_0^\infty\!d\eps\,\Re[v_1-i_\eps v_2]\,\Gamma(0,2\eps r),
\end{align}
which are, however, summed up to $\<\psi_j^\+(r)\sigma_3\psi_j(r)\>=0$ in both the subcritical and supercritical cases.
On the other hand, the lowest contribution to the angular current density $\<\psi_j^\+(r)\sigma_1\psi_j(r)\>$ is $O(1)$ and is evaluated as
\begin{align}
\mathrm{Fig.\,\ref{fig:polarization}} = \frac1\pi\int_0^\infty\!d\eps\,\Re[v_1-i_\eps v_2]\,e^{-2\eps r},
\end{align}
which leads to $\<\psi_j^\+(r)\sigma_1\psi_j(r)\>=\bar\gamma/(2g_1\pi r)$ in the subcritical case of Eq.~(\ref{eq:infrared}) but
\begin{align}
\<\psi_j^\+(r)\sigma_1\psi_j(r)\> 
= -\frac{\gamma}{2g_1\pi r}\int_0^\infty\!dz\,\Re\!\left[\tan\!\left(\gamma\ln\frac{r\Lambda_*}{z}\right)\right]e^{-z}
\end{align}
in the supercritical case of Eq.~(\ref{eq:supercritical}).
Therefore, while the angular current density has a power-law tail even for the subcritical Coulomb potential~\cite{Jackiw:2009}, its constant coefficient is turned into a log-periodic function of $r$ by the supercritical Coulomb potential.
Because only one partial-wave sector can be supercritical for $|g_0|\ll1$, the current density in the angular direction (\ref{eq:current}) including all partial waves is found to be
\begin{align}\label{eq:current_tail}
J_\theta(r) = -\frac{e\vF}{2\pi^2r^2}
\left\{\# - \frac{\gamma}{2g_1}\int_0^\infty\!dz\,\Re\!\left[\tan\!\left(\gamma\ln\frac{r\Lambda_*}{z}\right)\right]e^{-z}\right\},
\end{align}
where $\#$ is an unknown constant contributed by all subcritical sectors.

\section{Summary and conclusion}
In summary, the ABCD problem, i.e., massless Dirac fermions in a Coulomb potential accompanied by an Aharonov--Bohm solenoid [Eq.~(\ref{eq:hamiltonian})], was studied from a field-theoretic perspective.
To this end, we wrote down an effective action describing one partial-wave sector [Eq.~(\ref{eq:action})] and performed a perturbative renormalization group analysis enabled by suppressing the centrifugal barrier with the Aharonov--Bohm effect.
We showed that a scattering amplitude for reflection of particle at the potential center flows into an infrared fixed point [Eq.~(\ref{eq:infrared})] when the Coulomb potential is subcritical, while it exhibits the renormalization group limit cycle [Eq.~(\ref{eq:supercritical})] for the supercritical Coulomb potential revealing the emergence of discrete scale invariance.
Such a peculiar behavior is physically reflected not only in the geometric sequence of atomic collapse resonances [Eq.~(\ref{eq:resonance})] but also in the induced charge density [Eq.~(\ref{eq:charge_tail})] and current density [Eq.~(\ref{eq:current_tail})], both of which were found to have power-law tails with coefficients log-periodic with respect to the distance from the potential center.
Hopefully, our intriguing findings can in principle be realized by graphene experiments with charged impurities~\cite{Wang:2012,Wang:2013,Luican-Mayer:2014,Mao:2015}, where the induced charge and current densities are multiplied by four due to spin and valley degeneracy.
While they are all consistent with the previous prediction obtained by directly solving the Dirac equation~\cite{Nishida:2014}, our field-theoretic approach will be advantageous in incorporating the so-far-neglected electron-electron interaction~\cite{Biswas:2007}.

\acknowledgments
The author thanks Leonid S.~Levitov for valuable discussions at ISSP in June 2015 and for sharing his note on ``Atomic Collapse via Aharonov--Bohm Effect'' which led to this work.
This work was supported by JSPS KAKENHI Grants No.~JP15K17727 and No.~JP15H05855.

\onecolumngrid\appendix
\section{\label{sec:renormalization}Evaluation of diagrams in Fig.~\ref{fig:renormalization} for renormalization group equations}
The diagrams in Fig.~\ref{fig:renormalization} are logarithmically divergent at $\Lambda\to\infty$ and are evaluated as
\begin{align}
\mathrm{Fig.\,\ref{fig:renormalization}(a)}
&= \int_{\Lambda^{-1}}\!\frac{dr}{r}\,(g_0\sigma_0+g_1\sigma_1)
= (g_0\sigma_0+g_1\sigma_1)\ln\Lambda + \text{finite terms}, \\[2pt]
\mathrm{Fig.\,\ref{fig:renormalization}(b)}
+ \mathrm{Fig.\,\ref{fig:renormalization}(c)}
&= \sum_{a=0}^3\int_{\Lambda^{-1}}\!\frac{dr}{r}\,(g_0\sigma_0+g_1\sigma_1)G(\eps,r)v_a\sigma_a
+ \sum_{b=0}^3\int_{\Lambda^{-1}}\!\frac{dr}{r}\,v_b\sigma_bG(\eps,-r)(g_0\sigma_0+g_1\sigma_1) \notag\\
&= i_\eps g_0(v_0\sigma_0+v_1\sigma_1+v_2\sigma_2+v_3\sigma_3)\ln\Lambda
+ g_0(v_2\sigma_1-v_1\sigma_2)\ln\Lambda \notag\\
&\quad + i_\eps g_1(v_1\sigma_0+v_0\sigma_1)\ln\Lambda
+ g_1(v_2\sigma_0+v_0\sigma_2)\ln\Lambda + \text{finite terms}, \\[2pt]
\mathrm{Fig.\,\ref{fig:renormalization}(d)}
&= \sum_{a,b=0}^3\int_{\Lambda^{-1}}\!\frac{dr}{r}\,v_b\sigma_bG(\eps,-r)(g_0\sigma_0+g_1\sigma_1)G(\eps,r)v_a\sigma_a \notag\\
&= -\frac{g_1}{2}[(v_0^2+v_1^2-v_2^2-v_3^2)\sigma_1
+ 2v_1(v_0\sigma_0+v_2\sigma_2+v_3\sigma_3)]\ln\Lambda \notag\\
&\quad + i_\eps\frac{g_1}{2}[(v_0^2-v_1^2+v_2^2-v_3^2)\sigma_2
+ 2v_2(v_0\sigma_0+v_1\sigma_1+v_3\sigma_3)]\ln\Lambda + \text{finite terms}.
\end{align}
From the coefficients of $\sigma_{0,1,2,3}$, the renormalization group equations for $v_{0,1,2,3}$ are extracted to be
\begin{subequations}
\begin{align}
\frac{dv_0}{d\ln\Lambda} &= g_0 + i_\eps g_0v_0 + i_\eps g_1v_1 + g_1v_2 - g_1v_0v_1 + i_\eps g_1v_0v_2, \\
\frac{dv_1}{d\ln\Lambda} &= g_1 + i_\eps g_0v_1 + g_0v_2 + i_\eps g_1v_0 - \frac{g_1}{2}(v_0^2+v_1^2-v_2^2-v_3^2) + i_\eps g_1v_1v_2, \\
\frac{dv_2}{d\ln\Lambda} &= i_\eps g_0v_2 - g_0v_1 + g_1v_0 + i_\eps\frac{g_1}{2}(v_0^2-v_1^2+v_2^2-v_3^2) - g_1v_1v_2, \\
\frac{dv_3}{d\ln\Lambda} &= i_\eps g_0v_3 - g_1v_1v_3 + i_\eps g_1v_2v_3.
\end{align}
\end{subequations}
By combining these couplings to match with each element of $\sum_{a=0}^3v_a\sigma_a$, their renormalization group equations are simplified to
\begin{subequations}
\begin{align}
\frac{d(v_0\pm v_3)}{d\ln\Lambda} &= -[g_1(v_1-i_\eps v_2)-i_\eps g_0](v_0\pm v_3-i_\eps), \\
\frac{d(v_1+i_\eps v_2)}{d\ln\Lambda} &= -g_1(v_0+v_3-i_\eps)(v_0-v_3-i_\eps), \\
\frac{d(v_1-i_\eps v_2)}{d\ln\Lambda} &= -g_1(v_1-i_\eps v_2)^2 + 2i_\eps g_0(v_1-i_\eps v_2) + g_1.
\end{align}
\end{subequations}
As discussed in the main text below Eq.~(\ref{eq:propagator}), the complex coupling $v_1-i_\eps v_2$ represents a scattering amplitude for reflection of particle at the potential center, while the other couplings have no physical meaning in our semi-infinite one-dimensional problem.
To confirm the unitarity of solutions presented in Eqs.~(\ref{eq:subcritical}) and (\ref{eq:supercritical}), it is useful to derive the renormalization group equation for $|v_1-i_\eps v_2|^2$ from Eq.~(\ref{eq:running}), which is
\begin{align}
\frac{d|v_1-i_\eps v_2|^2}{d\ln\Lambda} = g_1[(v_1-i_\eps v_2)+(v_1-i_\eps v_2)^*](1-|v_1-i_\eps v_2|^2),
\end{align}
and thus $|v_1-i_\eps v_2|=1$ corresponds to its fixed point.

\section{\label{sec:polarization}Evaluation of diagrams in Figs.~\ref{fig:polarization} and \ref{fig:renormalization} for charge and current densities}
The diagrams in Figs.~\ref{fig:polarization} and \ref{fig:renormalization} contributing to $\<\psi_j^\+(r)\sigma_c\psi_j(r)\>$ are $O(1)$ and $O(g)$, respectively, and are expressed as
\begin{align}
\mathrm{Fig.\,\ref{fig:polarization}}
&= -\sum_{a=0}^3\int_{-\infty}^\infty\!\frac{d\eps}{2\pi}\,\Tr[\sigma_cG(\eps,r)v_a\sigma_aG(\eps,-r)], \\
\mathrm{Fig.\,\ref{fig:renormalization}(a)}
&= -\int_{-\infty}^\infty\!\frac{d\eps}{2\pi}\int_{\Lambda^{-1}}^\infty\!\frac{dr'}{r'}\,\Tr[\sigma_cG(\eps,r-r')(g_0\sigma_0+g_1\sigma_1)G(\eps,r'-r)], \\
\mathrm{Fig.\,\ref{fig:renormalization}(b)}
&= -\sum_{a=0}^3\int_{-\infty}^\infty\!\frac{d\eps}{2\pi}\int_{\Lambda^{-1}}^\infty\!\frac{dr'}{r'}\,\Tr[\sigma_cG(\eps,r-r')(g_0\sigma_0+g_1\sigma_1)G(\eps,r')v_a\sigma_aG(\eps,-r)], \\
\mathrm{Fig.\,\ref{fig:renormalization}(c)}
&= -\sum_{b=0}^3\int_{-\infty}^\infty\!\frac{d\eps}{2\pi}\int_{\Lambda^{-1}}^\infty\!\frac{dr'}{r'}\,\Tr[\sigma_cG(\eps,r)v_b\sigma_bG(\eps,-r')(g_0\sigma_0+g_1\sigma_1)G(\eps,r'-r)], \\
\mathrm{Fig.\,\ref{fig:renormalization}(d)}
&= -\sum_{a,b=0}^3\int_{-\infty}^\infty\!\frac{d\eps}{2\pi}\int_{\Lambda^{-1}}^\infty\!\frac{dr'}{r'}\,\Tr[\sigma_cG(\eps,r)v_b\sigma_bG(\eps,-r')(g_0\sigma_0+g_1\sigma_1)G(\eps,r')v_a\sigma_aG(\eps,-r)],
\end{align}
with $c=0$ for the charge density, $c=3$ for the radial current density, and $c=1$ for the angular current density.
While most of these expressions can be evaluated straightforwardly as presented in the main text, the part of Fig.~\ref{fig:renormalization}(a) for $c=0$ proportional to $g_0$ is tricky to evaluate as we elaborate here.

The tricky part is nothing short of the linear term of the following resummed expression:
\begin{align}
n_0(r) &= \int_{-\infty}^\infty\!\frac{d\eps}{2\pi}\,\Tr\!\left[\<r|\frac{1}{i\eps-\hat{H}_0-\hat{V}_0}-\frac{1}{i\eps-\hat{H}_0}|r\>\right] \notag\\
&= \int_{-\infty}^\infty\!\frac{d\eps}{2\pi}\int_{-\infty}^\infty\!dr'\,\Tr\!\left[\<r|\frac{1}{i\eps-\hat{H}_0-\hat{V}_0}|r'\>V_0(r')\<r'|\frac{1}{i\eps-\hat{H}_0}|r\>\right].
\end{align}
Here $\<r|\hat{H}_0|r'\>=-i\d_r\sigma_3\delta(r-r')$ is the kinetic operator and $\<r|\hat{V}_0|r'\>=V_0(r)\delta(r-r')$ is the potential operator with $V_0(r)=-(g_0/r)\theta(r-\Lambda^{-1})$ in our case.
The use of energy eigenfunctions $e^{ikr\sigma_3}$ for $\hat{H}_0$ and $e^{iKr\sigma_3-i\sigma_3\int_0^r\!dr''V_0(r'')}$ for $\hat{H}_0+\hat{V}_0$ leads to
\begin{align}
n_0(r) = \int_{-\infty}^\infty\!\frac{d\eps}{2\pi}\int_{-\infty}^\infty\!dr'\int_{-\infty}^\infty\!\frac{dK}{2\pi}\int_{-\infty}^\infty\!\frac{dk}{2\pi}\,\Tr\!\left[\frac{1}{i\eps-K}\frac{1}{i\eps-k}V_0(r')e^{i(K-k)(r-r')\sigma_3-i\sigma_3\int_{r'}^r\!dr''V_0(r'')}\right].
\end{align}
By integrating over $\eps$ and then $r'$, we obtain
\begin{align}
n_0(r) = 4\pi\int_{-\infty}^\infty\!\frac{dK}{2\pi}\int_{-\infty}^\infty\!\frac{dk}{2\pi}\,[\theta(-K)\theta(k)-\theta(K)\theta(-k)]\delta(K-k-V_0(r))
= -\frac{V_0(r)}{\pi},
\end{align}
which is presented in Eq.~(\ref{eq:density}) for $r>\Lambda^{-1}$.

\twocolumngrid


\begin{thebibliography}{99}

\bibitem{Bohr:1913}
N.~Bohr,
``On the constitution of atoms and molecules,''
Philos.\ Mag.\ \textbf{26}, 1-25 (1913);
Philos.\ Mag.\ \textbf{26}, 476-502 (1913);
Philos.\ Mag.\ \textbf{26}, 857-875 (1913).

\bibitem{Dirac:1928}
P.~A.~M.~Dirac,
``The quantum theory of the electron,''
Proc.\ R.\ Soc.\ London A \textbf{117}, 610-624 (1928);
Proc.\ R.\ Soc.\ London A \textbf{118}, 351-361 (1928).

\bibitem{Berestetskii:1982}
See, for example,
V.~B.~Berestetskii, L.~P.~Pitaevskii, and E.~M.~Lifshitz,
\textit{Quantum Electrodynamics}
(Butterworth-Heinemann, Oxford, 1982).

\bibitem{Pomeranchuk:1945}
I.~Ya.~Pomeranchuk and Y.~A.~Smorodinsky,
``On the energy levels of systems with $Z>137$,''
J.\ Phys.\ USSR \textbf{9}, 97-100 (1945).

\bibitem{Zel'dovich:1971}
Ya.~B.~Zel'dovich and V.~S.~Popov,
``Electronic structure of superheavy atoms,''
Sov.\ Phys.\ Usp.\ \textbf{14}, 673-694 (1972).

\bibitem{Case:1950}
K.~M.~Case,
``Singular potentials,''
Phys.\ Rev.\ \textbf{80}, 797-806 (1950).

\bibitem{Gitman:2012}
This procedure is also called the self-adjoint extension.
See, for example,
D.~M.~Gitman, I.~V.~Tyutin, and B.~L.~Voronov,
\textit{Self-Adjoint Extensions in Quantum Mechanics}
(Springer, Berlin, 2012).

\bibitem{Nishida:2014}
Y.~Nishida,
``Vacuum polarization of graphene with a supercritical Coulomb impurity: Low-energy universality and discrete scale invariance,''
Phys.\ Rev.\ B \textbf{90}, 165414 (2014).

\bibitem{Pereira:2007}
V.~M.~Pereira, J.~Nilsson, and A.~H.~Castro~Neto,
``Coulomb impurity problem in graphene,''
Phys.\ Rev.\ Lett.\ \textbf{99}, 166802 (2007).

\bibitem{Shytov:2007a}
A.~V.~Shytov, M.~I.~Katsnelson, and L.~S.~Levitov,
``Vacuum polarization and screening of supercritical impurities in graphene,''
Phys.\ Rev.\ Lett.\ \textbf{99}, 236801 (2007).

\bibitem{Shytov:2007b}
A.~V.~Shytov, M.~I.~Katsnelson, and L.~S.~Levitov,
``Atomic collapse and quasi-Rydberg states in graphene,''
Phys.\ Rev.\ Lett.\ \textbf{99}, 246802 (2007).

\bibitem{Wang:2012}
Y.~Wang, V.~W.~Brar, A.~V.~Shytov, Q.~Wu, W.~Regan, H.-Z.~Tsai, A.~Zettl, L.~S.~Levitov, and M.~F.~Crommie,
``Mapping Dirac quasiparticles near a single Coulomb impurity on graphene,''
Nat.\ Phys.\ \textbf{8}, 653-657 (2012).

\bibitem{Wang:2013}
Y.~Wang, D.~Wong, A.~V.~Shytov, V.~W.~Brar, S.~Choi, Q.~Wu, H.-Z.~Tsai, W.~Regan, A.~Zettl, R.~K.~Kawakami, S.~G.~Louie, L.~S.~Levitov, and M.~F.~Crommie,
``Observing atomic collapse resonances in artificial nuclei on graphene,''
Science \textbf{340}, 734-737 (2013).

\bibitem{Luican-Mayer:2014}
A.~Luican-Mayer, M.~Kharitonov, G.~Li, C.-P.~Lu, I.~Skachko, A.-M.~B.~Gon\c{c}alves, K.~Watanabe, T.~Taniguchi, and E.~Y.~Andrei,
``Screening charged impurities and lifting the orbital degeneracy in graphene by populating Landau levels,''
Phys.\ Rev.\ Lett.\ \textbf{112}, 036804 (2014).

\bibitem{Mao:2015}
J.~Mao, Y.~Jiang, D.~Moldovan, G.~Li, K.~Watanabe, T.~Taniguchi, M.~R.~Masir, F.~M.~Peeters, and E.~Y.~Andrei,
``Realization of a tunable artificial atom at a charged vacancy in graphene,''
Nat.\ Phys.\ \textbf{12}, 545-549 (2016).

\bibitem{Levitov:2015}
L.~S.~Levitov (private communication); see
\verb|http://www.mit.edu/~levitov/Lunchtalk_2012.pdf|.

\bibitem{Chakraborty:2011}
B.~Chakraborty, K.~S.~Gupta, and S.~Sen,
``Effect of topology on the critical charge in graphene,''
Phys.\ Rev.\ B \textbf{83}, 115412 (2011).

\bibitem{Kaplan:2009}
D.~B.~Kaplan, J.-W.~Lee, D.~T.~Son, and M.~A.~Stephanov,
``Conformality lost,''
Phys.\ Rev.\ D \textbf{80}, 125005 (2009).

\bibitem{Moroz:2010}
S.~Moroz and R.~Schmidt,
``Nonrelativistic inverse square potential, scale anomaly, and complex extension,''
Ann.\ Phys.\ (NY) \textbf{325}, 491-513 (2010).

\bibitem{Wilson:1971}
K.~G.~Wilson,
``Renormalization group and strong interactions,''
Phys.\ Rev.\ D \textbf{3}, 1818-1846 (1971).

\bibitem{Gorsky:2014}
A.~Gorsky and F.~Popov,
``Atomic collapse in graphene and cyclic renormalization group flow,''
Phys.\ Rev.\ D \textbf{89}, 061702(R) (2014).

\bibitem{Bulycheva:2014}
K.~M.~Bulycheva and A.~S.~Gorsky,
``Limit cycles in renormalization group dynamics,''
Phys.\ Usp.\ \textbf{57}, 171-182 (2014).

\bibitem{Terekhov:2008}
I.~S.~Terekhov, A.~I.~Milstein, V.~N.~Kotov, and O.~P.~Sushkov,
``Screening of Coulomb impurities in graphene,''
Phys.\ Rev.\ Lett.\ \textbf{100}, 076803 (2008).

\bibitem{Jackiw:2009}
R.~Jackiw, A.~I.~Milstein, S.-Y.~Pi, and I.~S.~Terekhov,
``Induced current and Aharonov--Bohm effect in graphene,''
Phys.\ Rev.\ B \textbf{80}, 033413 (2009).

\bibitem{Biswas:2007}
R.~R.~Biswas, S.~Sachdev, and D.~T.~Son,
``Coulomb impurity in graphene,''
Phys.~Rev.~B \textbf{76}, 205122 (2007).

\end{thebibliography}
\end{document}